\documentclass[aps,prl,twocolumn,longbibliography,superscriptaddress,floatfix,nobibnotes]{revtex4-2}

\usepackage{amsmath}
\usepackage{graphicx}
\usepackage{mathtools}
\usepackage{xcolor}
\usepackage{float}
\usepackage{bm}
\usepackage[normalem]{ulem}
\usepackage{dcolumn}

\begin{document}

\title{Large exchange bias from coupled antiferromagnetic order and glassy magnetic freezing in Co$_{0.36}$NbS$_2$}

\newcommand{\BGU}{Department of Physics, Ben-Gurion University of the Negev, Be'er-Sheva 84105, Israel}
\newcommand{\NRCN}{Physics Department, NRCN, P.O. Box 9001, Be'er Sheva, 84190, Israel}

\newcommand{\NHMFL}{National High Magnetic Field Laboratory, Tallahassee, Florida 32310, USA}

\author{Suman Mondal}
\affiliation{\BGU}

\author{Himadri Roy Dakua}
\affiliation{\BGU}

\author{Raz Luger} 
\affiliation{\BGU}

\author{Muntaser Naamneh} 
\affiliation{\BGU}

\author{Eran Maniv}
\email{Corresponding author : eranmaniv@bgu.ac.il}
\affiliation{\BGU}

\date{\today}

\begin{abstract}

Magnetization measurements on single-crystalline Co$_{0.36}$NbS$_2$ reveal antiferromagnetic order near $T_N \approx 26$ K and a lower-temperature history-dependent magnetic response consistent with glassy freezing of a subset of magnetic moments. After field cooling, $M(H)$ loops exhibit a sizable exchange-bias shift, reaching $|H_{\mathrm{EB}}| \approx 4.4$ kOe at 1.8 K for a 70 kOe field-cooling/sweep protocol. The training effect, field-sweep and cooling-field protocol dependence, and disappearance of the bias near a blocking temperature $T_B \approx 12$ K, well below $T_N$, support an interpretation in which frozen uncompensated moments are exchange-coupled to an antiferromagnetic matrix. These results suggest that antiferromagnetic-glassy-moment coupling may provide a broader route to exchange bias in off-stoichiometric magnetic NbS$_2$ intercalates.

\end{abstract}

\maketitle

\section{INTRODUCTION}

In exchange-biased (EB) systems, field cooling through a magnetic ordering temperature can produce a unidirectional anisotropy that shifts the hysteresis loop along the magnetic-field axis~\cite{bean,charles,meiklejohn}. Although EB was first established in conventional ferromagnet/antiferromagnet interfaces, similar loop shifts have since been observed in systems containing coupled ferromagnetic (FM), antiferromagnetic (AFM), ferrimagnetic, canted-AFM, and glassy magnetic components~\cite{maniv,ali,nayak,nayak2,sun,pradhan}. In bulk single-phase materials, EB is often associated with internal coupling between long-range magnetic order and disorder-induced frozen moments or short-range magnetic clusters~\cite{nayak,nayak2,sharma}. In such systems, the ordered magnetic background provides a pinning landscape, while the glassy component hosts uncompensated moments that can be biased by field cooling. A particularly relevant example is Fe$_{1/3+\delta}$NbS$_2$, where a giant low-temperature EB field of approximately 30 kOe was interpreted in terms of coupling between AFM order and disorder-induced glassy freezing~\cite{maniv}.

Intercalated transition-metal dichalcogenides (ITMDs) provide a natural platform for testing whether exchange bias from coupled AFM order and glassy freezing is a material-specific phenomenon or a broader consequence of magnetic disorder in off-stoichiometric intercalates. In 2H-NbS$_2$, intercalated 3d ions occupy the van der Waals gap, introduce local magnetic moments, and couple to the itinerant electronic states of the host. The balance between RKKY exchange, superexchange, spin-orbit coupling, and lattice-symmetry constraints makes the magnetic ground state highly sensitive to intercalant identity, concentration, and site occupancy~\cite{parkin1,parkin2,ghimire,regmi, mayoh, chopark, acharya}. Co$_{0.36}$NbS$_2$ is therefore a particularly useful test case: it is close to the commensurate $1/3$ intercalation, where long-range magnetic order is expected, but contains excess Co that can introduce occupational disorder and uncompensated frozen moments. Establishing exchange bias in this compound would show that the AFM-glassy-moment mechanism is not unique to Fe$_{1/3+\delta}$NbS$_2$, but may extend to Co-intercalated NbS$_2$ as well.

Here we show that overintercalated Co$_{0.36}$NbS$_2$ exhibits a sizable field-cooled EB effect associated with the coexistence of AFM order and glassy low-temperature magnetism. Magnetization measurements identify an AFM transition near $T_N \approx 26$ K, together with thermomagnetic irreversibility, remanent relaxation, and zero-field-cooled aging-memory effects consistent with glassy magnetic freezing. After field cooling, $M(H)$ loops show a clear EB shift reaching $|H_{\mathrm{EB}}| \approx 4.4$ kOe at 1.8 K. The training effect, field-sweep and cooling-field dependence, and disappearance of the bias near a blocking temperature $T_B \approx 12$ K, well below $T_N$, support a picture in which frozen uncompensated moments are exchange-coupled to an AFM matrix. Compared with Fe$_{1/3+\delta}$NbS$_2$, the bias is smaller but qualitatively similar, suggesting that AFM-glassy-moment coupling may be a broader route to EB in off-stoichiometric magnetic NbS$_2$ intercalates.

\begin{figure*}
    \centering
    \includegraphics[width=0.8\linewidth]{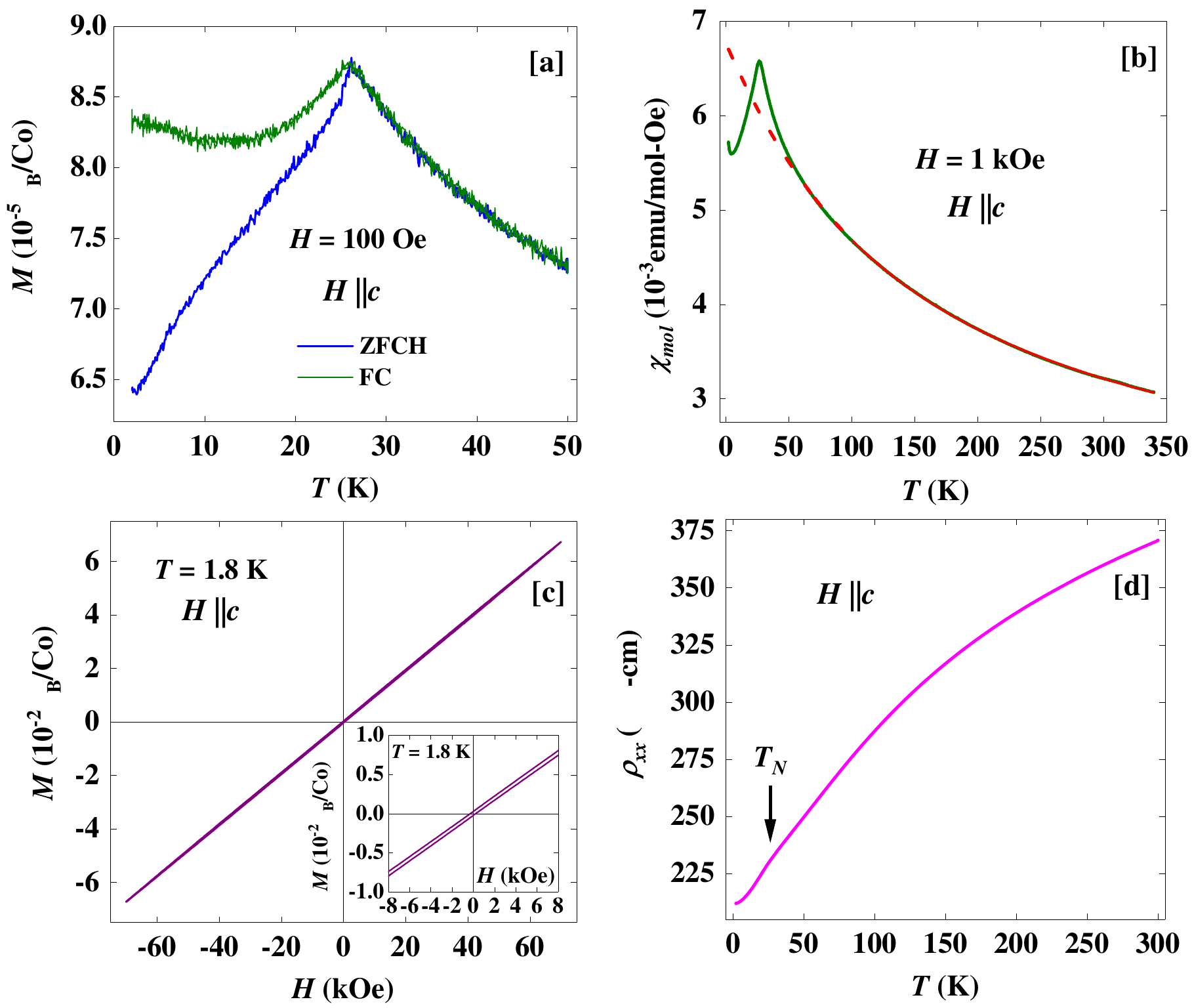}
     \caption{\textbf{Magnetic and transport characterization of Co$_{0.36}$NbS$_2$.} (a) Temperature dependence of the magnetization measured in zero-field-cooled heating (ZFCH) and field-cooled (FC) protocols for $H=100$ Oe applied along the $c$ axis. The cusp near $T_N\approx26$ K marks the AFM transition, while the ZFCH/FC bifurcation indicates thermomagnetic irreversibility. (b) Molar susceptibility, $\chi_{\mathrm{mol}}=M/H$, obtained after converting the measured magnetization to molar units. The red curve is a Curie--Weiss fit using $\chi=\chi_0+C/(T-\theta_P)$ over the range 100--340 K. (c) Zero-field-cooled $M(H)$ loop measured at 1.8 K in the field range $\pm70$ kOe. The inset enlarges the low-field region, showing a small hysteretic component centered near zero field. (d) Temperature dependence of the four-probe longitudinal resistivity $\rho_{xx}$, measured with $I\parallel ab$, $V_{\mathrm{meas}}\parallel I$, and an excitation current of $50~\mu\mathrm{A}$. The resistivity shows metallic behavior with a slope change near $T_N$.}
    \label{fig:FIG1}
    \end{figure*}   

\section{Results}   

\subsection{Antiferromagnetic Order and Magnetic Frustration}

Fig.~\ref{fig:FIG1}a shows the temperature-dependent magnetization measured in ZFCH and FC protocols for $H=100$ Oe applied along the $c$ axis. A cusp near $T_N\approx26$ K identifies the onset of AFM order. Upon further cooling, the ZFCH and FC curves gradually separate, indicating thermomagnetic irreversibility and the development of a history-dependent low-temperature magnetic component.

To estimate the dominant magnetic interaction scale, the high-temperature susceptibility was analyzed using a modified Curie-Weiss form, $\chi=\chi_0+C/(T-\theta_P)$, after converting the magnetization to molar susceptibility (Fig.~\ref{fig:FIG1}b). A fit over 100-340 K gives $\theta_P=-147.0\pm0.3$ K and $\mu_{\mathrm{eff}}\approx2.55~\mu_B/\mathrm{Co}$. The negative Curie--Weiss temperature indicates dominant AFM interactions, while the large separation between $|\theta_P|$ and $T_N$ gives a frustration factor $f=|\theta_P|/T_N\approx5.7$, comparable to values reported for other frustrated magnets~\cite{mohanty,miyata,kageyama,haley}. This reduced ordering temperature is consistent with moderate magnetic frustration and disorder in the overintercalated Co sublattice. The stability of the Curie-Weiss parameters against changes in fitting range is shown in Table S3 of the Supplementary Information.

The zero-field-cooled $M(H)$ loop measured at 1.8 K (Fig.~\ref{fig:FIG1}c) is dominated by an approximately linear high-field response, consistent with an AFM background, but also contains a small low-field hysteretic component. The small magnitude of this hysteresis suggests that the low-temperature loop opening is associated with frozen uncompensated moments rather than a conventional FM phase. Together with the ZFCH/FC irreversibility, this motivates the field-cooled exchange-bias measurements discussed below.

The temperature-dependent longitudinal resistivity $\rho_{xx}(T)$ (Fig.~\ref{fig:FIG1}d) shows metallic behavior over the measured temperature range, with a clear slope change near $T_N$. This anomaly suggests coupling between charge transport and the onset of magnetic order. The relatively low residual resistivity ratio, $\mathrm{RRR}=\rho_{xx}(300~\mathrm{K})/\rho_{xx}(2~\mathrm{K})=1.76$, is consistent with appreciable disorder, likely associated with off-stoichiometric Co occupation~\cite{maniv2021antiferromagnetic}. Additional magnetoresistance and Hall measurements, including the absence of a clear anomalous Hall contribution within the measured field range, are shown in Fig.~S3 of the Supplementary Information.
   
\subsection{Exchange Bias and Training Effect}

\begin{figure*}[t!]
    \includegraphics[width=1\linewidth]{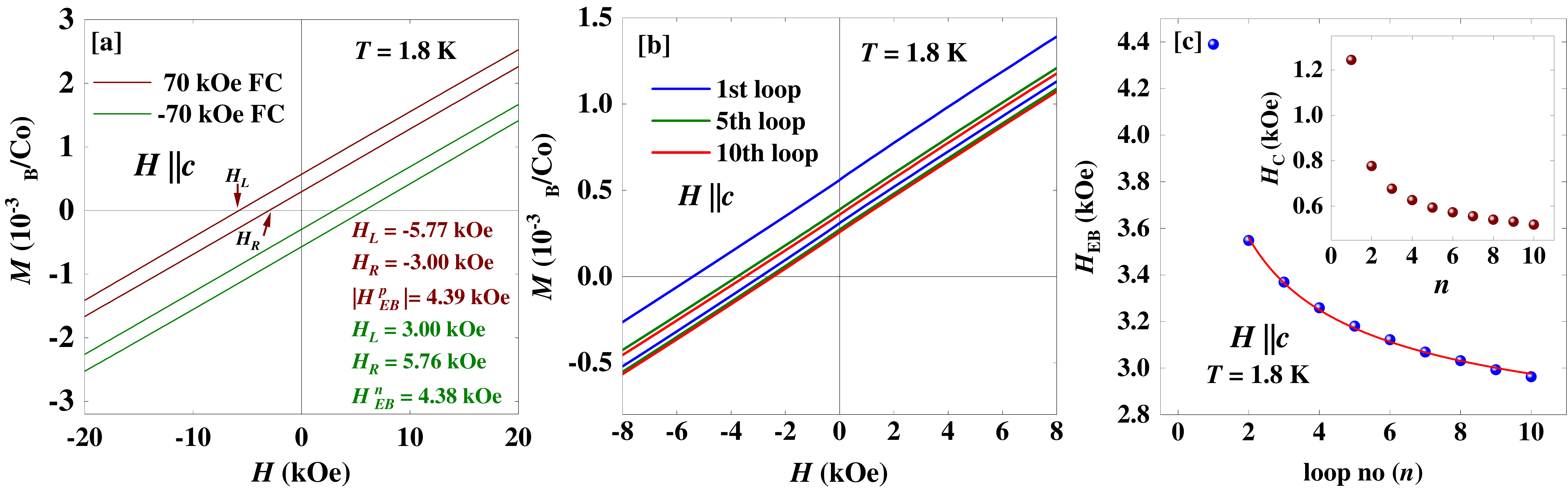}
    \caption{\textbf{Field-cooled exchange bias and training effect.} (a) Field-cooled $M(H)$ loops measured at 1.8 K after cooling in $+70$ kOe and $-70$ kOe with $H\parallel c$. Cooling in positive (negative) field shifts the loop toward negative (positive) field, demonstrating a field-cooled exchange-bias response. The lower- and higher-field zero crossings, $H_L$ and $H_R$, are indicated. (b) Training-effect measurement after field cooling in $+70$ kOe. The first, fifth, and tenth hysteresis loops are shown. (c) Exchange-bias field $|H_{\mathrm{EB}}|$ as a function of cycle number $n$. The solid red line is a fit for $n\ge2$ to Binek's phenomenological training relation. The inset shows the corresponding coercive field $H_C$.}
     \label{fig:FIG2}
\end{figure*}

The lower- and higher-field zero crossings of each hysteresis loop are denoted by $H_L$ and $H_R$, respectively. The EB field and coercive field are defined as \[H_{\mathrm{EB}}=\frac{H_L+H_R}{2}, \qquad H_C=\frac{H_R-H_L}{2}.\] In the following, we plot the EB magnitude $|H_{\mathrm{EB}}|$. Fig.~\ref{fig:FIG2}a shows field-cooled $M(H)$ loops measured at 1.8 K after cooling in $+70$ kOe and $-70$ kOe. Cooling in positive field shifts the loop toward negative field, while cooling in negative field reverses the shift direction. The extracted EB magnitudes are $|H_{\mathrm{EB}}|\approx4.39$ kOe and $4.38$ kOe for positive and negative field cooling, respectively, demonstrating a sizable field-cooled EB response.

\par

A characteristic feature of exchange-biased magnetic systems is the training effect, in which the loop shift decreases upon repeated field cycling. Fig.~\ref{fig:FIG2}b shows selected hysteresis loops measured at 1.8 K after cooling in $+70$ kOe and cycling the field between $\pm70$ kOe. The EB shift decreases most strongly between the first and second cycles and then evolves more gradually with increasing cycle number, while remaining finite after ten cycles. This behavior is consistent with previous reports of training in exchange-biased systems~\cite{bansal,zheng}. The cycle dependence of $H_{\mathrm{EB}}$ is commonly described by Binek's phenomenological training relation~\cite{binek},
  
\begin{eqnarray}
H_{\mathrm{EB}}(n)-H_{\mathrm{EB}}^{\infty}=\frac{k}{\sqrt{n}} \nonumber\\
\label{eqn1}
\end{eqnarray}

Here, $H_{\mathrm{EB}}^{\infty}$ is the exchange-bias field extrapolated to large cycle number and $k$ is a constant. Fitting the data for $n\ge2$ gives $H_{\mathrm{EB}}^{\infty}=2.50\pm0.01$ kOe and $k=1.49\pm0.03$ kOe. The fit for $n\ge2$ captures the gradual part of the training effect after the large irreversible change between the first and second loops. This behavior is consistent with partial reconfiguration of weakly pinned uncompensated moments under repeated field cycling, while a more strongly pinned component remains stable and maintains a finite EB shift~\cite{zabel}. Branch-dependent changes during training are shown in  Fig.~S4 of the Supplementary Information. Such asymmetry further indicates that repeated cycling modifies the field-cooled pinned-spin configuration, rather than producing a simple symmetric reduction of the hysteresis width.

\begin{figure*}
\centering
\includegraphics[width=1\linewidth]{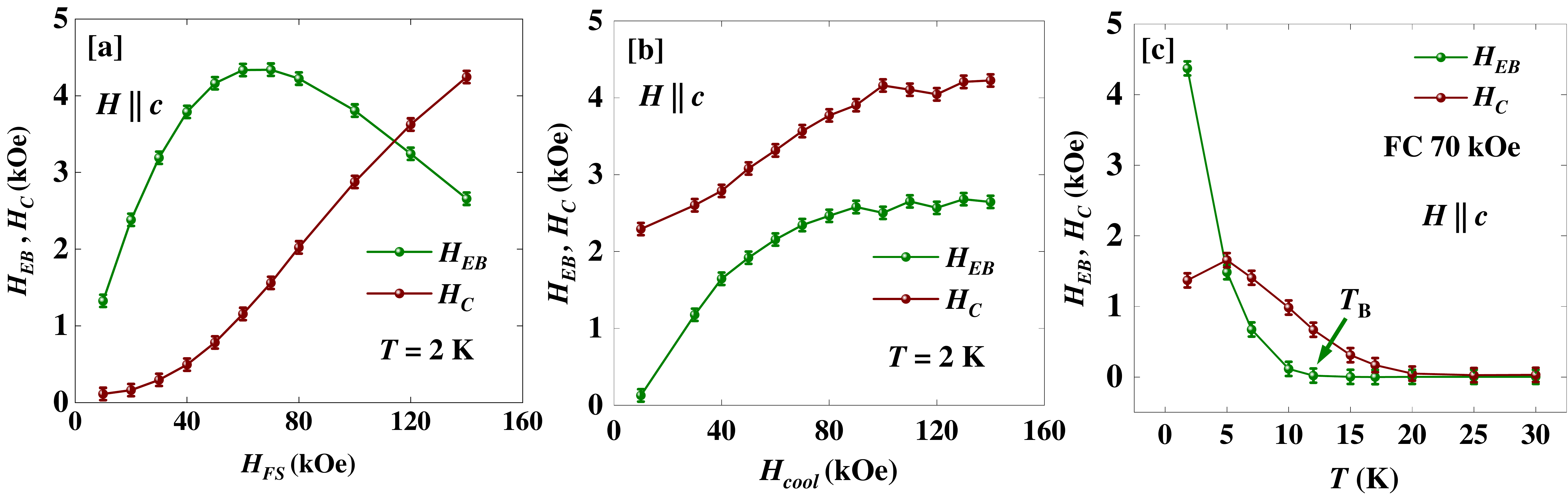}
\caption{\textbf{Field-protocol and temperature dependence of the EB response.}
(a) EB field $|H_{\mathrm{EB}}|$ and coercive field $H_C$ as a function of the field-sweep range $H_{\mathrm{FS}}$, measured at 2 K with $H\parallel c$. In this protocol, the cooling field was set equal to the sweep field, $H_{\mathrm{cool}}=H_{\mathrm{FS}}$. (b) Cooling-field dependence of $|H_{\mathrm{EB}}|$ and $H_C$ at 2 K, measured using a fixed maximum field-sweep range of $\pm140$ kOe. (c) Temperature dependence of $|H_{\mathrm{EB}}|$ and $H_C$ after field cooling in 70 kOe and measuring loops with a $\pm70$ kOe sweep range. The EB vanishes near the blocking temperature $T_B\approx12$ K, while $H_C$ becomes small on approaching $T_N$. Error bars represent the uncertainty estimated from the magnetic-field step size.}
\label{fig:FIG3}
\end{figure*}

We next examine how the EB response depends on magnetic-field protocol. Two complementary measurements were performed. In the first, the cooling field and the maximum field-sweep range were varied together, $H_{\mathrm{cool}}=H_{\mathrm{FS}}$ (Fig.~\ref{fig:FIG3}a). In the second, the maximum sweep range was fixed at $\pm140$ kOe while only the cooling field was varied (Fig.~\ref{fig:FIG3}b). This separation is important because the measured loop shift can depend not only on the field used during cooling, but also on the field range used to trace the hysteresis loop.

In the $H_{\mathrm{cool}}=H_{\mathrm{FS}}$ protocol, $|H_{\mathrm{EB}}|$ initially increases with field, reaches a maximum value of approximately $4.39$ kOe at $H_{\mathrm{FS}}=70$ kOe, and then decreases at larger sweep fields (Fig.~\ref{fig:FIG3}a). The initial increase likely reflects stronger field alignment of the uncompensated pinned component during cooling, whereas the reduction at larger fields suggests that the larger measurement sweep can partially depin or reconfigure weakly pinned moments.

To separate cooling-field effects from sweep-range effects, we also measured loops using a fixed maximum field sweep of $\pm140$ kOe while varying only $H_{\mathrm{cool}}$ (Fig.~\ref{fig:FIG3}b). In this fixed-sweep protocol, $|H_{\mathrm{EB}}|$ increases with cooling field and tends to saturate at high $H_{\mathrm{cool}}$. Thus, the cooling field tunes the field-cooled pinned-spin configuration, but the high-field decrease observed in Fig.~\ref{fig:FIG3}a is associated with the increasing sweep range rather than with cooling field alone.

The temperature dependence further constrains the exchange-bias mechanism. After cooling in 70 kOe, $|H_{\mathrm{EB}}|$ decreases monotonically with increasing temperature and vanishes near a blocking temperature $T_B\approx12$ K (Fig.~\ref{fig:FIG3}c). This temperature lies well below the AFM ordering temperature, $T_N\approx26$ K. Therefore, the disappearance of exchange bias at $T_B<T_N$ indicates that long-range AFM order alone is insufficient to produce the loop shift; a frozen uncompensated component is also required. The coercive field decreases with temperature and becomes small on approaching $T_N$, supporting the assignment of the low-temperature hysteresis to frozen glassy moments coupled to the AFM background.

Because EB-like loop shifts can sometimes arise from unsaturated minor loops, we also considered this possibility. In the present data, the sign reversal of the loop shift upon reversing the cooling-field direction, the training effect, and the systematic field-protocol dependence argue against a simple minor-loop artifact. The raw field-cooled loops used to extract Fig.~\ref{fig:FIG3} are shown in Figs.~S6, S7, and S8 of the Supplementary Information.

\begin{figure*}[htbp!]
\centering
\includegraphics[width=1\linewidth]{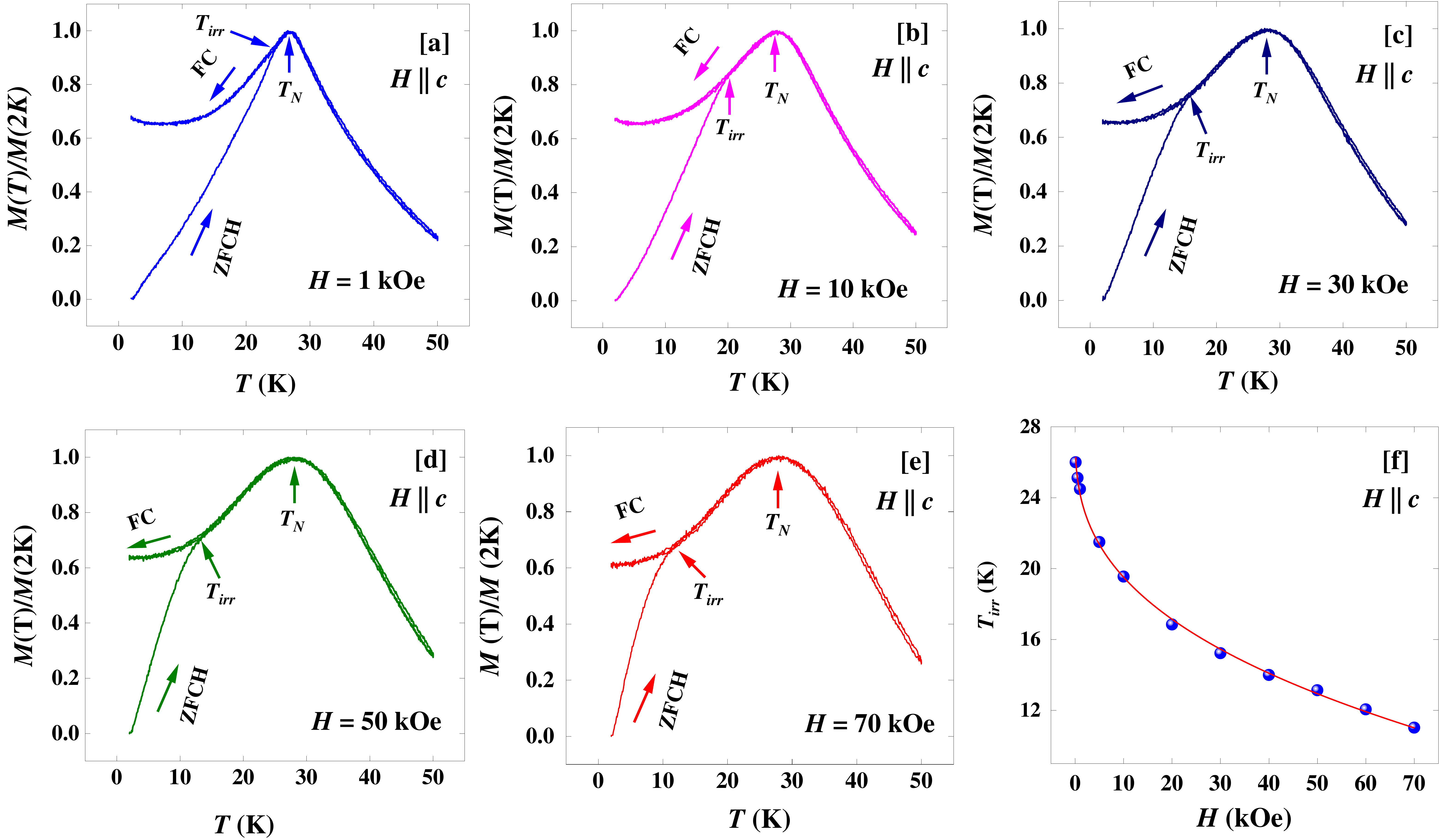}
\caption{\textbf{Field dependence of magnetic irreversibility.}
(a)-(e) Normalized magnetization, $M(T)/M_{\mathrm{FC}}(2~\mathrm{K})$, measured in ZFCH and FC protocols for different applied fields with $H\parallel c$. The cusp near $T_N$ marks the AFM transition, while $T_{\mathrm{irr}}$ is defined operationally as the temperature at which the ZFCH and FC curves bifurcate.
(f) Field dependence of $T_{\mathrm{irr}}$. The solid red line is a fit to Eq. (2), giving $\phi=5.54\pm0.29$.}
\label{fig:FIG4}
\end{figure*}
           
\subsection{Glassy Magnetic Freezing}

Previous work on Fe$_{1/3+\delta}$NbS$_2$ showed that a giant exchange-bias response can arise when AFM order coexists with disorder-induced glassy freezing~\cite{maniv}. In that system, NMR measurements revealed a distorted AFM state, while magnetic measurements showed glassy irreversibility and training effects, supporting a picture of frozen uncompensated moments coupled to an AFM background. This provides a useful reference point for Co$_{0.36}$NbS$_2$, where excess Co can similarly introduce occupational disorder within an otherwise AFM-ordered intercalant sublattice. Motivated by this analogy, we next examine whether the low-temperature irreversibility in Co$_{0.36}$NbS$_2$ is consistent with a field-suppressed glassy freezing component.

The low-temperature ZFCH/FC bifurcation in Co$_{0.36}$NbS$_2$ suggests the onset of a history-dependent magnetic state. To examine the field dependence of this irreversibility, we measured $M(T)$ in ZFCH and FC protocols under several applied fields with $H\parallel c$ (Fig.~\ref{fig:FIG4}a-e). We define $T_{\mathrm{irr}}$ operationally as the temperature at which the ZFCH and FC curves begin to separate. With increasing magnetic field, $T_{\mathrm{irr}}$ shifts to lower temperature, decreasing from approximately 21.5 K at 5 kOe to approximately 11 K at 70 kOe. This field suppression of thermomagnetic irreversibility is consistent with a glassy or frozen low-temperature magnetic component. Phenomenologically, the field dependence of $T_{\mathrm{irr}}$ can be described by
\begin{eqnarray}
T_{\mathrm{irr}}(H)=T_{\mathrm{irr}}(0)\left(1-BH^{2/\phi}\right) \nonumber\\ 
\label{eqn2}
\end{eqnarray}
where $T_{\mathrm{irr}}(0)$ is the zero-field extrapolated irreversibility temperature, $B$ is a constant, and $\phi$ parameterizes the field dependence. In mean-field descriptions of spin glasses, the Almeida-Thouless form corresponds to $\phi=3$, whereas the Gabay-Toulouse form expected in the low-anisotropy limit corresponds to $\phi=1$. The data are well described by Eq.~\ref{eqn2}, yielding $\phi=5.54\pm0.29$ (Fig.~\ref{fig:FIG4}f). This value is far from the low-anisotropy GT limit and is broadly consistent with a strongly anisotropic glassy-freezing scenario. Comparable exponents have been reported in other glassy magnetic systems, including DyNi, Cr$_{0.5}$Fe$_{0.5}$Ga, and Mn$_{0.25}$Ti$_{1.1}$S$_2$~\cite{dyni,pallab,MnTiS}. Consistent with this picture, thermoremanent magnetization measurements show slow relaxation at 1.8 K and a gradual loss of the remanent component on warming, as shown in Fig.~S5 of the Supplementary Information. 


\begin{figure}
    \centering
    \includegraphics[width=1\linewidth]{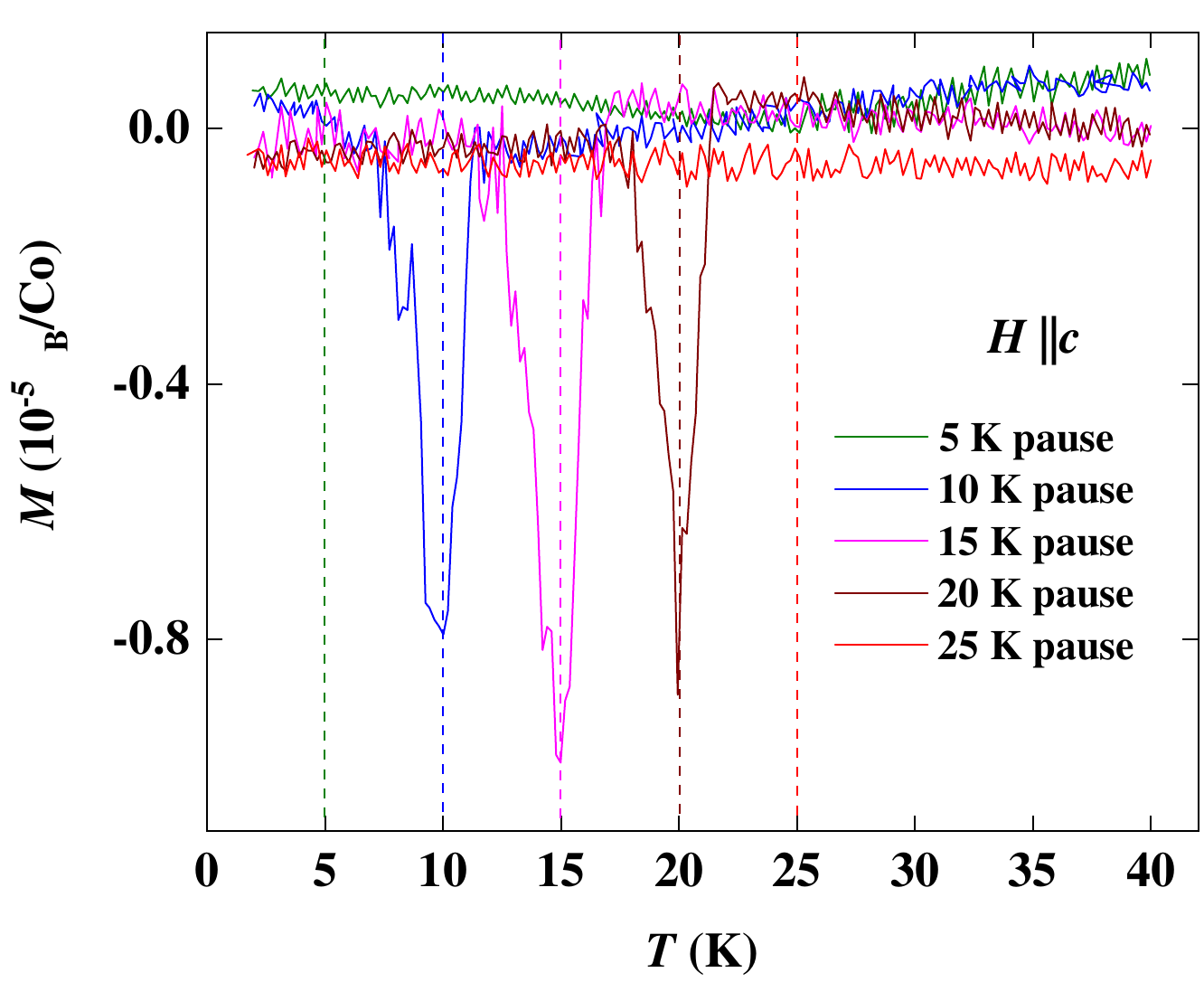}
    \caption{\textbf{ZFC aging-memory response of Co$_{0.36}$NbS$_2$.} The sample was cooled in zero field with a one-hour waiting time at selected temperatures $T_w$, then cooled to base temperature. Magnetization was measured on warming in $H=1$ kOe and compared with a reference curve measured without a waiting step. The plotted difference, $\Delta M(T)=M_{\mathrm{aged}}(T)-M_{\mathrm{ref}}(T)$, shows memory dips near the waiting temperatures, most clearly for $T_w=10$, 15, and 20 K. The dips indicate that the system retains memory of the aging temperature, consistent with slow glassy magnetic dynamics.}
    \label{fig:FIG5}
\end{figure}

To further test the history dependence of the low-temperature magnetic state, we performed ZFC aging-memory measurements. In this protocol, the sample was cooled in zero field with an intermediate waiting step at a selected temperature $T_w$, then cooled to base temperature. The magnetization was subsequently measured during warming in a small applied field and compared with a reference curve measured without a waiting step. The difference $\Delta M(T)=M_{\mathrm{aged}}(T)-M_{\mathrm{ref}}(T)$ is shown in Fig.~\ref{fig:FIG5}, while the corresponding aged and reference warming curves are shown in Fig.~S9 of the Supplementary Information. For waiting temperatures of 10, 15, and 20 K, $\Delta M(T)$ exhibits memory dips near $T_w$, indicating that the system retains a record of the aging temperature.

Such aging-memory behavior is a hallmark of slow glassy dynamics: during the waiting step, part of the magnetic configuration relaxes toward a metastable state characteristic of that temperature, and this configuration is partially recovered upon reheating~\cite{mydosh,vincent}. The observation of memory dips within the AFM state supports the coexistence of long-range AFM order with a history-dependent frozen magnetic component. Notably, the memory response can extend over a broader temperature range than the exchange-bias blocking temperature, indicating that the onset of slow magnetic dynamics and the formation of a robust unidirectional bias are related but not identical. EB requires sufficiently pinned uncompensated moments coupled to the AFM background, whereas thermomagnetic irreversibility and aging can persist even when the field-cooled loop shift is already strongly reduced. The observation of aging-memory behavior provides an additional parallel with Fe$_{1/3+\delta}$NbS$_2$, where remanent relaxation and memory effects were used to identify disorder-induced glassy freezing~\cite{maniv}. 

\begin{figure}
    \centering
    \includegraphics[width=1\linewidth]{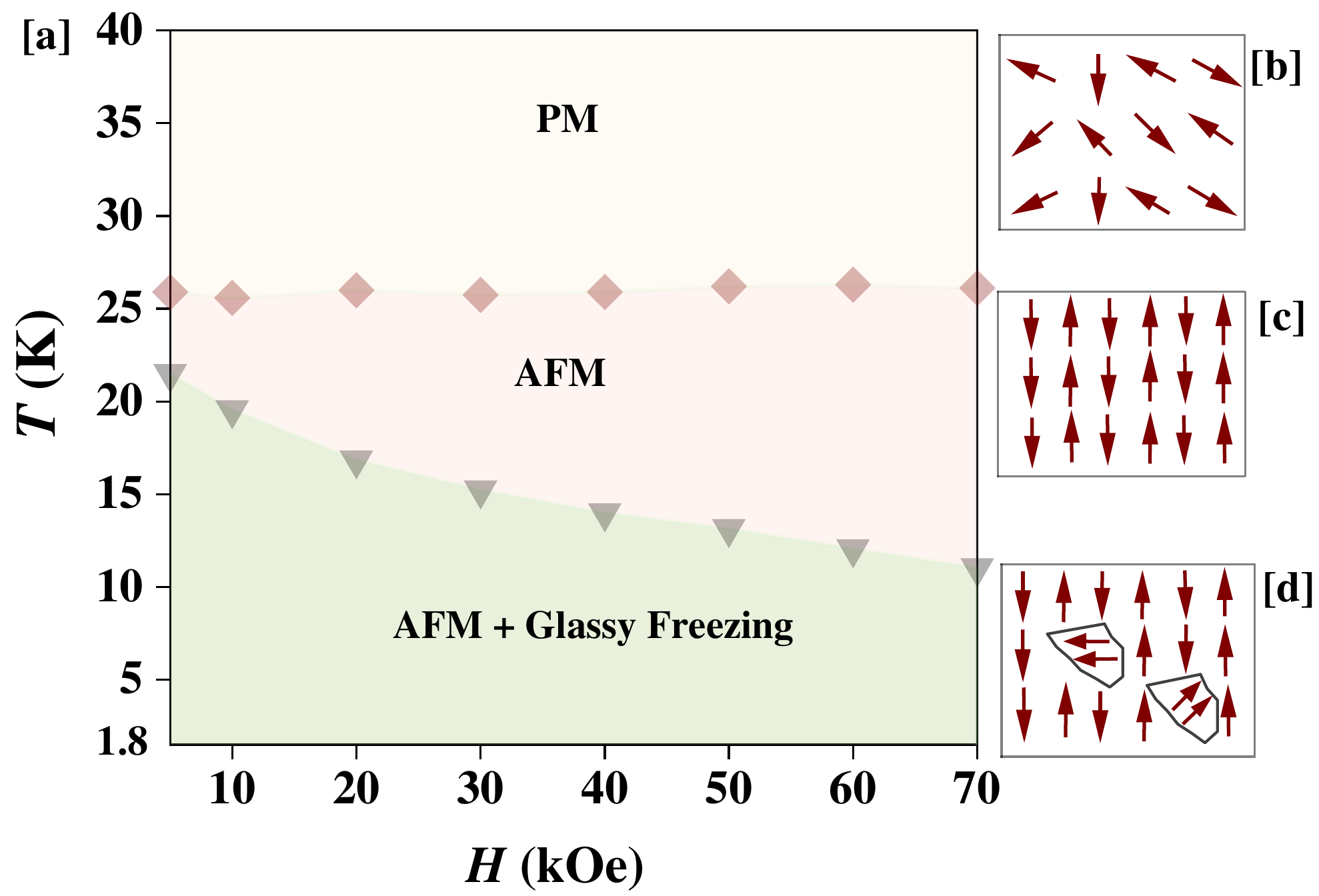}
    \caption{\textbf{Magnetic phase diagram and schematic exchange-bias mechanism.} (a) Magnetic field-temperature phase diagram for Co$_{0.36}$NbS$_2$ with $H\parallel c$. The upper boundary marks the AFM transition near $T_N$, while the lower boundary follows the irreversibility temperature $T_{\mathrm{irr}}$ extracted from the ZFCH/FC bifurcation. Symbols indicate experimentally determined $T_{\mathrm{irr}}$ values; shaded regions are guides to the eye. (b)-(d) Phenomenological schematic of the proposed field-cooling mechanism. Frozen uncompensated moments associated with Co-rich or disordered regions couple to the AFM background and generate a unidirectional anisotropy after field cooling. The schematic does not imply direct microscopic imaging of clusters or domains.}
    \label{fig:FIG6}
\end{figure}

Combining the ZFCH/FC irreversibility, remanent relaxation, aging-memory response, and EB blocking temperature yields the phase diagram summarized in Fig.~\ref{fig:FIG6}a. The data indicate three regimes: a high-temperature paramagnetic regime, an intermediate AFM regime below $T_N$, and a lower-temperature regime in which AFM order coexists with history-dependent glassy freezing. The exchange bias appears only in the lowest-temperature regime, where frozen uncompensated moments are sufficiently pinned to generate a unidirectional anisotropy after field cooling.

\section{Discussion}

The combined magnetic and transport measurements identify Co$_{0.36}$NbS$_2$ as a moderately frustrated antiferromagnet with a history-dependent low-temperature magnetic response. The negative Curie--Weiss temperature and the frustration factor $f\approx5.7$ indicate dominant AFM interactions whose ordering temperature is suppressed relative to the interaction scale. Below $T_N\approx26$ K, the emergence of ZFCH/FC irreversibility, remanent relaxation, and aging-memory behavior points to a frozen magnetic component coexisting with the AFM background. The field-induced suppression of $T_{\mathrm{irr}}$, summarized in Fig.~\ref{fig:FIG6}a, is consistent with glassy magnetic freezing. Because this assignment is based on DC magnetization and remanence, frequency-dependent AC susceptibility would be required to determine whether the frozen component corresponds to a canonical spin glass or to a more general glassy cluster state.

The measured Co content exceeds the ideal $1/3$ intercalation, suggesting partial occupation of additional intercalation sites and the presence of occupational disorder in the Co sublattice~\cite{acharya}. Such disorder can introduce random exchange pathways and locally uncompensated moments embedded within the AFM background. Neutron scattering would be required to determine how the magnetic structure of overintercalated Co$_{0.36}$NbS$_2$ differs from that of stoichiometric Co$_{1/3}$NbS$_2$~\cite{mangelsen,chopark}, while local probes such as MFM, Lorentz TEM, or magnetic imaging would be needed to directly identify frozen clusters or domain textures. Within the phenomenological picture summarized in Fig.~\ref{fig:FIG6}, frozen uncompensated moments associated with Co-rich or disordered regions are exchange-coupled to the AFM matrix. Field cooling then selects a metastable pinned configuration, producing a unidirectional anisotropy and hence an EB shift.

The disappearance of EB near $T_B\approx12$ K, well below $T_N$, shows that AFM order alone is insufficient to generate the loop shift. Instead, the bias appears only when the low-temperature frozen component is sufficiently pinned. The proximity between $T_B$ and the high-field irreversibility scale extracted from ZFCH/FC measurements suggests that the EB is tied to the onset of robust glassy freezing. The persistence of a finite bias under large sweep fields further supports strong AFM-mediated pinning of the frozen uncompensated component, although the precise microscopic pinning mechanism remains to be established.

It is useful to compare this behavior with the reported exchange bias in low-intercalated Co$_{0.30}$NbS$_2$, where a loop shift of approximately 2 kOe was observed at 2 K for an applied field of 90 kOe~\cite{mangelsen}. The larger bias observed here in Co$_{0.36}$NbS$_2$ may suggest that overintercalation-induced occupational disorder is particularly effective in generating pinned glassy moments. However, because the measurements were performed on different compositions and protocols, a systematic composition-dependent study would be required to establish how the EB magnitude evolves across Co$_x$NbS$_2$.

The present results place Co$_{0.36}$NbS$_2$ in direct analogy with Fe$_{1/3+\delta}$NbS$_2$, where off-stoichiometry tunes the balance between AFM order and disorder-induced glassy freezing. In Co$_{0.36}$NbS$_2$, the exchange-bias scale is smaller than in Fe$_{1/3+\delta}$NbS$_2$, but the phenomenology is similar: both systems show AFM order, low-temperature glassy irreversibility, training, and a field-cooled loop shift. The reduced bias in the Co compound may reflect differences in local moment size, magnetic anisotropy, AFM stiffness, disorder landscape, or the density and pinning strength of frozen uncompensated moments. Determining which of these factors dominates will require microscopic probes of the magnetic structure and local frozen texture. Nevertheless, the observation of sizable EB in Co$_{0.36}$NbS$_2$ supports the broader idea that AFM-glassy-moment coupling can generate exchange bias in off-stoichiometric magnetic NbS$_2$ intercalates.

\section{Conclusions} 

In summary, Co$_{0.36}$NbS$_2$ exhibits AFM order near $T_N\approx26$ K together with a low-temperature history-dependent magnetic response consistent with glassy freezing. Field-cooled hysteresis loops show a sizable EB shift, reaching $|H_{\mathrm{EB}}|\approx4.4$ kOe at 1.8 K for the 70 kOe field-cooling/sweep protocol. The training effect, field-sweep and cooling-field dependence, aging-memory response, and disappearance of the bias near $T_B\approx12$ K, well below $T_N$, support an interpretation in which frozen uncompensated moments are exchange-coupled to an AFM matrix. These results establish overintercalated Co$_{0.36}$NbS$_2$ as a platform for studying exchange bias generated internally by the coexistence of AFM order, disorder, and slow glassy magnetic dynamics.

\section{Methods}

Single crystals of Co$_{0.36}$NbS$_2$ were grown by chemical vapor transport. Appropriate amounts of Co, Nb, and S were first reacted in an evacuated quartz ampoule at $900^\circ$C to form a polycrystalline precursor. Single crystals were then grown from this precursor using iodine as the transport agent in a temperature gradient from $940^\circ$C to $860^\circ$C for seven days. The crystal structure and composition were characterized by single-crystal X-ray diffraction and SEM-EDX. EDX analysis gave an average Co content of $x=0.36\pm0.01$ in Co$_x$NbS$_2$. Further structural and compositional characterization is provided in the Supplementary Information (Figs.~S1 and S2, and Table~S2).

Magnetization measurements were performed using a Quantum Design MPMS3 SQUID magnetometer and the vibrating-sample magnetometer option of a Quantum Design PPMS Dynacool. The magnetic field was applied along the crystallographic $c$ axis. The crystal mass used for magnetization measurements was approximately $3.8$ mg. Temperature-dependent magnetization was measured in zero-field-cooled heating (ZFCH) and field-cooled (FC) protocols. Exchange-bias loops were measured after cooling from above $T_N$ to the target temperature in a specified cooling field. The field-sweep range was chosen according to the measurement protocol and is specified in the corresponding figures and Supplementary Information. Unless stated otherwise, the sample was warmed above $T_N$ in zero field between successive field-cooling protocols. Detailed field-cooling and field-sweep sequences are summarized in Table S1 of the Supplementary Information.

Electrical transport measurements were performed in a Quantum Design PPMS using mechanically exfoliated flakes contacted in a four-probe geometry. The current was applied in the $ab$ plane, and the longitudinal voltage was measured parallel to the current direction. The temperature-dependent longitudinal resistivity was measured using an excitation current of $50~\mu\mathrm{A}$.

\section{Data availability}
All data supporting the findings of this work are included within the manuscript. All other relevant data are available from the corresponding author upon request.

\section{Code availability}
The analysis code used to process the magnetization and transport data is available from the corresponding author upon request.

\section{References}
\bibliographystyle{naturemag}
\bibliography{ref}

\section{Acknowledgements}
Work by S.M., H.R.D., R.L. and E.M. was performed with support from the European Research Council grant no. ERC-101117478, the Israeli Science Foundation grant no. ISF-885/23 and the PAZY foundation grant no. 412/23. 

\section{Author contributions}
H.R.D. and S.M.  performed the crystal growth, magnetization measurements, transport measurements, and data analysis. R.L., and M.N. contributed to experimental measurements and analysis. E.M. conceived and supervised the project. S.M. and E.M. wrote the manuscript with input from all authors.

\section{Competing interests}
The authors declare no competing interests.

\end{document}